\documentclass[a4paper,11pt]{report}

\usepackage{a4wide}
\usepackage{graphicx}

\def\la{\mathrel{\mathchoice {\vcenter{\offinterlineskip\halign{\hfil$\displaystyle##$\hfil\cr<\cr\sim\cr}}}{\vcenter{\offinterlineskip\halign{\hfil$\textstyle##$\hfil\cr<\cr\sim\cr}}}{\vcenter{\offinterlineskip\halign{\hfil$\scriptstyle##$\hfil\cr<\cr\sim\cr}}}{\vcenter{\offinterlineskip\halign{\hfil$\scriptscriptstyle##$\hfil\cr<\cr\sim\cr}}}}}
\def\ga{\mathrel{\mathchoice {\vcenter{\offinterlineskip\halign{\hfil$\displaystyle##$\hfil\cr>\cr\sim\cr}}}{\vcenter{\offinterlineskip\halign{\hfil$\textstyle##$\hfil\cr>\cr\sim\cr}}}{\vcenter{\offinterlineskip\halign{\hfil$\scriptstyle##$\hfil\cr>\cr\sim\cr}}}{\vcenter{\offinterlineskip\halign{\hfil$\scriptscriptstyle##$\hfil\cr>\cr\sim\cr}}}}}

\begin{document}

\setlength{\parindent}{7mm}

\noindent
{\LARGE\bf The Water Circuit of the Plants}

\vspace{1mm}

\noindent
{\Large\bf Do Plants have Hearts ?}

\vspace{5mm}

\noindent
Wolfgang Kundt and Eva Gruber

\vspace{5mm}

\noindent
W. Kundt, Institut f\"ur Astrophysik der Universit\"at Bonn, Auf dem H\"ugel 71, D-53121 Bonn, e-mail: wkundt@astro.uni-bonn.de , Fax: 0228-73-3672.
Eva Gruber, Emil-Fischer-Gymnasium, Euskirchen.

\vspace{1cm}

Abstract: There is a correspondence between the circulation of blood in all higher animals and the circulation of sap in all higher plants - up to heights h of $\la$ 140 m - through the xylem and phloem vessels. Plants suck in water from the soil, osmotically through the roothair zone, and subsequently lift it osmotically again, and by capillary suction (via their buds, leaves, and fruits) into their crowns. In between happens a reverse osmosis - the endodermis jump - realized by two layers of subcellular mechanical pumps in the endodermis walls which are powered by ATP, or in addition by two analogous layers of such pumps in the exodermis. The thus established root pressure helps forcing the absorbed ground water upward, through the whole plant, and often out again, in the form of guttation, or exudation.

\vspace{1cm}
{\Large\bf 1. No Plant without Water}
\vspace{5mm}

We do not know a plant that could grow without water. Extreme cases of modest needs of water are fungi as well as all desert plants, but even such creatures decay when the air moisture drops permanently below some threshold, of order 60\%.

   Water is primarily required for the transport of materials, both from the ambient soil towards the solar factories in the leaves - or even branches: Pfanz et al (2002) - and from the leaves towards all the domains of growth, above and below ground. Water from the soil provides the indispensable anorganic nutrients; it is conducted upward through the xylem vessels to the leaves, in which photosynthesis converts carbon dioxide (from the ambient air) into energy-rich hydrocarbons, mainly into starch, with nitrogen- and phosphor-containing intermediate products. After the upward transport, water is again required for taking the photosynthesis products from the leaves through the phloem to all sites of growth, in the buds, branches, and stems, roots, blooms, and fruits. Water must therefore permanently circulate through every plant, as sap with dissolved materials of varying concentrations. Moreover, water is the basic substance of the fluid inside cells, the cytoplast, as well as the fluid between cells, the apoplast, and is also an indispensable building material of the cell membranes and walls as well as all other hardware of plants, like carbohydrates, lipids, proteins, and nucleic acids.

   The problem of how plants manage to lift their water from the soil to their leaves has been with biologists at least since 1726, when S. Hales discussed it in the first edition of the Proceedings of the Royal Society of London. So much has been thought about it that present-day scientists - including eminent referees - often believe the problem has been solved long ago, or that there was no problem at all. Mentioned are osmotic suction and capillary forces, transpiration and coherence. Why have they never been applied by civilization, to provide water for an elevated hut, or apartment? Osmotic and capillary forces can perform work once only (for given inhomogeneities), up to the state of maximal dilution or wettening, respectively. They can compensate transpiration losses. But they cannot propel a circulation, or provide extra supply (from the soil, or from some other reservoir), unless continually restored. Such tasks require pumps. Like animals, plants must perform work when raising their building materials against gravity, or transporting them from one reservoir to another. 

   Holding and raising long threads of water (inside the xylem tubes) requires their coherence, which prevents them from tearing. But coherence is not a sufficient property for raising: every rope is coherent, without being able to perform work. The coherence-tension theory of the rising of water in plants - which assumes that the leaves carry (and even raise) the threads of water below them, and which is still often defended by leading biologists until today - was proposed back in 1894 by Dixon \& Joly. But it ignores, among others, that the leaves (with their capillary vessels) are not rigid enough to carry tree-long liquid threads; they would be crushed under the load, which would enhance their weight by a factor of $\ga 10^{-2}$ h/l, of at least 10, for the tallest trees, where h is the height of the tree, l the length of a leaf, and the factor $10^{-2}$ estimates the relative water weight of the leaf. Besides, much of the lifting of water to the crowns of trees happens at night, in the absence of solar energy.

   Often overlooked in this connection is the fact that all higher plants possess subcellular mechanical pumps in their roothair zone, in the plasmodesmata of the outer and inner toroidal endodermis wall (which separates the cortex from the central cylinder), with which they establish the well-known root pressure $p$ which can split rock, bend fences, and lift houses, at many times atmospheric pressure, typically $p \la 10$ bar. Root pressure can be used by a plant to force absorbed water up into its leaves and even beyond, during guttation, preferentially at night. It also manifests itself by the `bleeding' (exudation) of cut branches, or stems, of alder, birch, chestnut, grapevine, kiwi, maple, oak, or walnut, mainly in early spring. 

   Big trees lift a ton of water daily into their crowns, smaller plants correspondingly less. Such lifting of water requires work: the sap must be forced up against gravity. Corresponding work is performed in the cities by stations which pump fresh water into the network of supplying pipes, or water towers. In higher animals, the heart takes care of circulating the blood through their body. For plants, this supply problem is not answered in the biological textbooks, at least not satisfactorily so. Here we update, and improve the approaches taken by Kundt and by Kundt \& Robnik in 1998, and in 2004.

\vspace{1cm}
{\Large\bf 2. How far are we from a consensus?}
\vspace{5mm}

There are three quite different ways in which water (or any other liquid) can be stored in long vertical tubes inside a tall building, or tree:

   (1) Without any additional device, a vertical cylindrical tube filled with water must only carry its own weight. But at the same time, it must hold the pressure of the contained water, which grows by 1 bar for every 10 m of height, reaching 14 bar for the tallest trees! If a small hole is punched through the tube near the ground, water will shoot out at speeds reaching $(2p/\rho)^{1/2} \la 190$ km/h, against experience. This mode of containment is therefore unlikely to be realized in tall trees.

   (2) In order to avoid the high sideways pressure of a heavy water column, an architect can try and attach the water to the ceiling of the vertical cylindrical tube, i.e. hang it at its top via surface tensions (exerted by capillaries). This is the assumption made by the coherence-tension theory; it replaces the high pressures near the ground by equally high tensions near the top. It is defended e.g. in Steudle (1995, 2001), and in Tyree (2003), and dismissed in Zimmermann et al (2004). Apart from the problems mentioned above, this mechanism is thought to be highly unstable to cavitation followed by tearing of the water threads. Other possible instabilities are expected due to mechanical damage of the xylem under its enormous tensions, and due to reduced robustness of the tree during storms (when its crown has to carry, in excess, the weight of all its xylem water). And: The `overlapping double saw-cuts' experiment proves that a tree does not die when all its xylem vessels are severed (Preston 1952, Zimmermann et al 2004), giving the theory a deadly blow. Even worse: one of us (EG) has witnessed a young maple tree whose stem was cut all the way around its periphery some ten years ago, which has continued growing ever since. 

   (3) A third way to store water in high trees is similar to the way in which high apartment houses are constructed, in which the weight of every compartment is transferred directly to the carrying walls. For water columns, this transfer of their weight can be made osmotically, by segmenting the xylem tubes via (inclined) porous plates across the tracheids, as detailed in the next section. In this way, the osmotic pressure $\pi$ grows linearly with the height $z$, in step with the gravitational potential $\rho gz$, the xylem pressure $p$ remains (almost) constant, of order 1 bar, and the weight of the water column is transferred to its container, the xylem vessel; cf. Zimmermann et al (1993).  

   Historically, the problem of the functioning of plants, and of the ascent of water in them has been predominantly approached via specific measurements rather than via physical constraints imposed by optimal functioning. But measurements on living plants are difficult, and measurements on dead plants can be misleading, so that misinterpretations are abundant. For instance, measurements of the xylem pressure $p$ with the Scholander-Hammel pressure chamber have yielded very different results from measurements with the R\"ade-Zimmermann (hollow-needle) pressure probe and its multiple variants, often by confounding pressure $p$ with osmotic pressure $\pi$(Salisbury \& Ross 1992 p.59, Zimmermann et al 2004, Tyree 1997). (Note that Scholander's ``pressure chamber'' is routinely called ``pressure bomb'' because two of its predecessor versions - made from glass by Henry Dixon - happened to explode). Other examples of hard-to-disentangle experiments are Hacke \& Sperry's (2003) refilling tests on certain dried plants of claimed negative xylem-sap pressure - a claim which we find hard to believe - and Schwenke \& Wagner's (1992) exudation experiments on specially treated excised (!) maize roots which show periodic (hourly) and spiky (sub-minute) rate fluctuations. Are we sufficiently educated to correctly interpret such complex observations? Progress of fundamental insight usually derives from simple, basic phenomena and/or considerations.

     In early days, the ascent of water in plants was explained by transpiration: solar energy is used to pull the ground water up into the crowns (Hales 1726, B\"ohm 1893). In modern textbooks, transpiration is still supported by deriving a huge (negative) water potential (of the atmosphere, forgetting to multiply the chemical potential by the density contrast between steam and liquid, of order $10^{-3}$ (e.g. L\"uttge \& Higinbotham 1979 p.286, Mohr \& Schopfer 1992 p.51, Salisbury \& Ross 1992 pp. 46, 104, L\"uttge et al 1999 p. 411). Such explanations often ignore that transpired water is (not gained but) lost for the plant, like for animals. And unlike often stated, the rate of transpiration (plus secretion) is comparable for plants and for animals, when compared with the circulating volume, or the retained water in the body: Plants have a much larger surface area than animals, but they transpire through only some 1\% of it, through the pores of the leaves whenever they open to take in carbon dioxide; most of the surface is sealed off by waxes (Matthews 2006). Transpiration is an unavoidable loss mechanism (of water). Several plants show even `reverse transpiration', i.e. a takeup of water through their leaves (Zimmermann et al 2004). On the other hand, transpiring causes an increasing osmolarity of the drying organs, and less wettened capillaries, so that its losses are replenished from below as long as water reserves are available. In this way, transpiration reactivates their pulling, and helps lifting the solutes of the soil water to the crown; it serves as an additional motor during epochs of enhanced growth. 

    What about storage mode (2), the hooking up of threads of water in the crowns by surface tension called `coherence-tension' mechanism ? This mode was first proposed in 1894 by Dixon \& Joly, and later resumed and supported by Pickard (1981), Milburn (1996), Steudle (1995, 2001), Tyree (1997, 2003), Tyree \& Zimmermann (2002), Woodward (2004), and many others. Is it feasible? It probably owes a lot of its present acceptance to the work by Pickard who has called it ``well understood'' and having ``stood the test of the time'' yet admits that ``residua of misunderstanding persist'', that ``additional understanding would be helpful'', ``fracturing is a major unsolved problem'', ``measurements of flow rate...fail to agree...with the predictions'', and that it ``is to some of us equivalent to believing in ropes of sand''. At the same time, he mentions a ``discredited root pressure'', ``the role of...osmosis...(that has) not been as well appreciated as might have been hoped'', and ``gradients of osmolarity which can be measured along a trunk or stem (which) do accord more readily with pull from above than with push from below''. In our understanding, Pickard paves the road to an osmotic interpretation of the ascent of sap, and to mode (3) of its support, and we agree with Zimmermann et al (2004) that ``stable negative pressures of the order of megapascals...belong in the realm of science fiction''. Dipl.-Ing Karlheinz Hahn (1993) speaks of a ``wissenschafts-historischer Irrweg''.

    We thus arrive at storage mode (3) of roughly uniform pressures $p$ throughout a (tall) plant, realized by osmotic segmenting. This idea goes back to Pfeffer (1881) and Sachs (1887) and is partially supported by Morizet \& Robelin (1972), Zholkevich (1991), Canny (1995,1998), Kundt (1998), and Zimmermann et al (2004). Even the recent measurements by Koch et al (2004) conform with it if the readings of their pressure chamber are properly interpreted (as osmotic pressures, instead of negative pressures: Salisbury \& Ross 1992 p.59). Osmotic gradients propel the sap, helped by the indispensable root pressure, in particular for small plants. This message of the present article cannot be found in full clarity in the quoted work, in part because of problems with the definition, and role of the water potential $\psi$, cf. Canny (1999): Water flows always downhill w.r.t. $\psi$, independently of how it is confined. Canny appreciates the problems with the coherence-tension theory, and replaces it by a ``Compensating Pressure'' theory for which it is not clear what parts of a plant perform which necessary work; the physical mechanisms he has in mind remain obscure. Similarly, Zimmermann et al (2004) propose a ``Multy-Force or Watergate'' theory without identifying the multiple forces. Their Marangoni effect - a modified surface tension - should be incorporated into the water potential, into its matric-potential term $\tau$, but its functioning is not clear. Again, they face unnecessary problems with $\psi$, which are evidenced by their use of terms like ``turgor pressure'' (=$p$), ``solute-reflecting barriers'' (without ``reflections''), ``reflection coefficients'' (which do not apply to equilibrium situations), which obscure their revolutionary findings of almost atmospheric xylem pressures throughout. We now turn to the major building blocks of the osmotic explanation.      

\vspace{1cm}
{\Large\bf 3. Water Currents}
\vspace{5mm}

Water flows downhill: from sites of higher gravity potential $\rho gz$ (for mass density $\rho$) to sites of lower potential, where $g$ is the gravity acceleration, and $z$ is the height above some zero level, more generally the distance from the center of Earth. Rivers prove it.

   If, however, water is enclosed in a system of pipes, it may occasionally flow uphill, namely in the direction of increasing height $z$. If, moreover, its osmotic concentration varies along a pipe, the osmotic suction must be taken care of as well: Dissolved materials attract water, in proportion to the number density of their constituent particles, independently of whether or not the latter are electrically neutral or charged, i.e. are molecules or ions. Intuitively, osmotic suction can be visualized as minus the pressure exerted by the gas of its enclosed, constituent particles which are regularly reflected by its walls and thereby exert a one-sided pressure, the osmotic pressure $\pi$, which can take values almost as large as $10^3$ times atmospheric pressure (kbar) for suitable solutes. This internal overpressure acts like a suction on the outside water. The adequate physical description then introduces the (preferred) water potential
\begin{equation}
 \psi := \rho gz + p - \pi - \tau
\end{equation}                               		 
which generalizes the gravity potential $\rho gz$ to systems of pipes filled inhomogeneously with aqueous solutions, and which is described more thoroughly in the appendix `water potential'; (the capillary potential, or `matric´ potential $\tau$ will be explained soon). With this definition, water flows always `downhill' in terms of $\psi$, i.e. from sites of higher potential to sites of lower potential, cf. Pickard (1981 p.217). In thermodynamics, $\psi$ is known as the (Gibbsian) free enthalpy, or as the chemical potential divided by the volume per particle (e.g. Nultsch 1991, who confusingly writes $\psi_p$ for $p$, $\psi_\pi$ for $\pi$, and subtracts from $\psi$ its value for pure water, which would destroy the crucial `flowing downhill' property).

   This insight has already revealed to us one of the secrets of the water supply in trees: Is the sap to rise, its osmotic concentration (osmolarity) in the crown must be higher than in its roots. If, on the other hand, the sap is to stream from the leaves towards the fruits, the latter must offer the higher concentration. A plant can achieve this, e.g., by converting starch into sugar (with a vastly higher number density of its molecules), or by secreting Ca ions in the form of calcium-oxalate crystals (in the vacuoles of the leaves: Braune et al 1979). Quantitatively, a mono-ionic concentration of 0.0446 mole/l exerts a tension of 1 bar, strong enough to pull water to a height of 10 m, cf. Fig.1b. Its osmotic pressure equals the atmospheric gas pressure.

   But reality is even more complex: Hydrophilic wall material exerts additional adhesive forces onto the sap whose importance grows with an increasing ratio of surface area over volume, i.e. with an increasing thinning of the tubes. Such capillary forces are known since centuries and help the leaves attract water; they act as long as there are unwettened internal surfaces. Strongest are these forces for imbibing seeds, whence their name `imbibition' (or `matric') suction; they are described by an additional potential, $\tau$, equal in sign to $\pi$, which can likewise take values as large as kbar in extreme cases. For terrestrial gravity, water in a hydrophilic tube of radius $r$ rises to a height of 15m($\mu$m/$r$), i.e. to a height of 15m for $r = 1 \mu$m, which increases as $r^{-1}$ with decreasing radius $r$, cf. Fig. 1a. But for decreasing $r$, there is a natural tradeoff between the height and the rate of its supply. And, as argued above, the leaves with the thinnest capillaries are not rigid enough to hold tree-long water columns.

   So how do tall trees manage to supply their crowns with soil water, up to heights $h$ of $\la 140$ m, for which the hydrostatic pressure varies by $p \la 14$ bar between top and ground level? No architect of high towers would hook the movable outfit to the roof. As mentioned above, there are the stability problems (of the leaves, the threads, and the stems), and the enormous tensions which the enclosing - even hydrophobic (!: Zimmermann et al 2004) - xylem vessels would have to take, reaching -13 bar and more near the tops of tall trees. Zimmermann et al's pressure probe has not detected them (1993, 2004); they arose due to misinterpretations of the Scholander pressure chamber. They would force trees to form unnecessarily heavy xylem vessels with increasing height. 

   Instead, the most economic solution of the water supply problem for tall trees is an almost constant water potential, where the increase with height of  $\rho gz$ is compensated by an equal increase of $\pi$ (in Equ.(1), for vanishing $\tau$ along the stem), so that the pressure $p$ remains constant, near 1 bar: piercing xylem vessels does not cause small im- or explosions. The higher a tree, the larger must be $\pi$ near its top (Koch et al 2004). Conifers (gymnosperms) approximate this solution ($\psi$ = const) via inclined walls with bordered pits across the tracheids once every few meters, serving as safety valves which are forced open during fresh supply from below, and maintain the necessary small upward jumps in osmolarity, whereas angiosperms (vascular plants) realize it via perforation plates which check the downward diffusive spreading of the solutes by a faster upward flow speed through a reduced cross section of the xylem vessels. In both realizations, the weight of the water columns is quasi continuously transferred to the wooden framework of the stem, so that their pressure stays near atmospheric.

   So far, we have described the static equilibrium, achieved by osmotic forces with a continual increase of the osmolarity with height. But how to propel the water? We offer two answers: (i) As long as the roots provide enough ground water, an organ that needs it should enhance its osmotic concentration, and tap the nearest xylem vessel. (ii) Another way to propel the xylem water is to enhance the root pressure, which forces new soil water towards the stem and up. More in detail, soil water is attracted osmotically, through the cortex of the primary roots in the roothair zone, see Fig. 2, and thereby enriched osmotically to pressures of $\la 10$ bar, comparable to the osmolarity of physiological salt solution - 6.23 bar - at the inner edge of the cortex. (Desert plants can even achieve root pressures of $\la 60$ bar). Here the attracted water would stagnate, and the leaves in the crown could not pull, would not the high osmolarity be lowered back to a level comparable to that of the soil: a reverse osmosis is required, and known to take place, as the endodermis jump: Ursprung \& Blum (1921), also Strasburger et al (1971). The sap in the xylem vessels of the roots tends to be almost tasteless (or somewhat bitter, due to salts of magnesium), i.e. highly diluted. Only in the crown does the osmotic pressure rise again to several bar, on average by 1 bar per 10 m, and the taste gets bitter, (to be contrasted by the tastelessness of guttation water).

   This reverse osmosis through the (cylindrical) endodermis walls, on transition from the cortex to the central cylinder, in two steps in series, reduces the concentration of the soil water back to near its original value, and allows the leaves in the crown to exert a second osmotic pull. It maintains the high pressure of the sap, in transition to the central cylinder, so that the sap is forced towards the stem and at the same time, the young roots can penetrate hard ground. Even though this reverse osmosis is textbook knowledge since many decades, under the name of `endodermis jump', the conclusion has apparently not been drawn elsewhere that the (mechanical) pumps in the endodermis perform the work necessary for both a second osmotic pull, and a pushing of the water crownward. I.e. the pumps serve as the hearts of the plants, without whom the plants would violate the second law of thermodynamics: no reverse osmosis without work.

    Some confusion about the ``endodermis jump'' has arisen through the claim in L\"uttge \& Higinbotham (1979, p.324) that ``apart from the Casparian strip in their radial walls, the cells of the endodermis represent nothing unique'', and that ``endodermis cells do not have the cytological characteristics of...cell types involved in solute pumping''. Indeed, no transfer cells, or ion pumps are meant to be involved. Yet the endodermis cells are not only unique by their (lignified) Casparian strips, but also by their $10^3$ plasmodesmata per toroidal wall (e.g. Braune et al 1979, L\"osch 2001, Schreiber et al 1994); and as argued in the last section, their pumps act as pure-water pumps, realized by their periodically squeezed desmotubules, which reduce the high osmolarity of the innermost cortex layer back to that of the ground water. If not, there would be a jump in solute flow from the ground to the xylem vessels which would give rise to a stagnation of the transport within the filling time of a cortex layer, less than a minute. (Selective retentions of ions must be less than $\epsilon = d/vt = 10^{-5}/t_6$ for a primary-root lifetime $t$ in units of $10^6$s, $t_6 := t/10^6$s = $t/10$d, where $\epsilon$ denotes the retained fraction by number). We have no reason for doubting the results of Ursprung \& Blum (1921).

\vspace{1cm}
{\Large\bf 4. Root Pressure}
\vspace{5mm}

The presence of (mechanical) water pumps in the roots of plants is not only required by the second law; it could alternatively have been concluded from the well known phenomena of root pressure, and exudation, which are not only observed when big trees raise their ambient soil, or bend obstacles, but become even more manifest when plants are wounded, or decapitated: Water pours out for days, from a wounded branch, stem, or stump, against the (holding) forces of osmosis and imbibition! Root pressure gets likewise manifest via guttation, observed even at the bottom of rain forests, in the dark, at 100\% air moisture, against osmotic suction. Guttation - the squeezing out of diluted water from the edges of leaves - can be seen daily in e.g. banana (plants), cabbage, grass, kiwi, lady's mantle rose (=Frauenmantel), strawberry, see Figs. 3, 4. Both exudation and guttation are opposite in sign to osmotic and capillary forces. They require an excess pressure inside a plant, not to be established by suction, which can amount to 6 bar in the tomato, 10 bar in grass stalks, or even 60 bar in certain desert plants. It has been well studied since many decades, for various plants (White et al 1958). This excess pressure allows plants to penetrate into the ground, split rock, lift concrete plates, and push away obstacles. Shoots from coconuts pierce their tough enclosing shells! 

   Root pressure shows a daily rhythm, and a yearly rhythm: the pumps only work when required, in an unspectacular way (Kramer \& Kozlowski 1979). Reassuringly for our interpretation, plants growing on salty soil, or in salt water, must work harder for reducing the osmolarity; such halophytes - and also many other plants (like rice, and mais) - have a second cylindrical array of plasmodesmal pumps, equally equipped with girdling Caspary strips, in the exodermis, right inside (beyond) the rhizodermis (Zimmermann et al 1992, Salisbury \& Ross 1992). They can reduce the potential of the entering water twice, in series. 

   Biologists tend to talk of `turgor' when they mean `pressure'. A container pumped up with a gas, or liquid gets tighter and smoother with increasing pressure, as long as it does not burst; and so do cells of plants. Turgor is the Latin word for pressure; it therefore does not make sense to talk of `turgor pressure'. Cells are stiffened by the pressure of their liquid (cytoplast), similar to balloons, tires, or water beds. 

   When talking of forces, it can be wise to evaluate the corresponding energy balance, for a more thorough understanding. A force $F$ acting along a differential path $dx$ performs the differential work $dW = F dx$. Correspondingly, a work per area $\Delta w = \int \rho g h dz$ must be performed in order to raise a liquid column of mass density $\rho$, height $h$, through a vertical distance $\int dz$ against gravity $g$. More generally, the water potential $\psi$ implies that the transport of a liquid column through a distance $dx$ requires the work per area
\begin{equation}
\Delta w = \int \psi dx .		
\end{equation}
In particular, (dilution) work is performed when water enters a volume of higher osmolarity, like sap rising in a high plant with a sufficient osmotic gradient. This gradient can be set up by the chemistry of the leaves (helped by the parenchym cells?), and by transpiration. In a steady state, however, and in the absence of transpiration, the driving pressure gradient is provided by the mechanical pumps at the root tips, in the roothair zone, which force the ground water (slowly) crownward. 

   Whichever way, raising a ton of water ($M$) per day to a height $h$ of 100 m requires an energy $W$ of $10^6$ Joule = 17 kWatt min , which equals the unscreened solar flux $A \int S dt$ throughout a quarter of an hour onto an area $A$ of 1 m$^2$: $W = M g h = 17 \mathrm{kWatt min} = A \int S dt$ (for above values). Clearly, photosynthesis can provide several 100 times this energy per summer day (for realistic $A$ and $\int dt$); not only animals can work. And water rises in plants not only during sunny daytime, but also at night, during dry epochs, and in winter, before the leaves have unfolded. It also rises at the bottom of rain forests - manifested by guttation - where transpiration is prevented by 100\% air moisture and almost complete darkness, as well as in plants kept locked up in closed glass containers for a whole season. In all such thirsty intervals, plants depend uniquely on their root pumps.    

\vspace{1cm}
{\Large\bf 5. The Water Pumps of the Plants}
\vspace{5mm}

We have argued that plants cannot do without water pumps, or hearts. Where and how are the pumps realized? Figure 2 summarizes our current understanding, which differs in detail from the proposals made earlier by Kundt \& coauthors, guided by new insight at the experimental front: Water intake happens mainly in the roothair zone of the (young) primary root tips, Fig.\,5, the latter of lengths between a mm and several cm, between the newly forming front part (dividing meristem) and the succeeding, somewhat older secondary root whose periphery is already corked and/or woodened. This location, at the very tips of the roots, is the obvious site for the pumps if the pressurized ground water is to be forced through the whole plant: through the thin and thick roots, stem and branches, all the way up to the leaves.      

   As a rule, these sucking primary root segments tend to be active during a few days to weeks only, with exceptions up to a year, helped by the unicellular root hairs which enlarge the surface area for ground water intake. Moreover, most plants profit during water intake from the `woodwide web', i.e. from extended networks of mycorrhicae which function as the merchants of the soil, trading nourishing ground water in the cortex for the products of photosynthesis (Read 1997). In any case, the osmotic pressure of the incoming solution grows in several steps of $\la$1 bar, from one cell layer to the next, from the exodermis to the endodermis, and with it the pressure $p$ (Zimmermann et al 1993). What happens beyond? 

   Beyond the cortex, Ursprung \& Blum (1921) have shown that the high osmotic pressure is abruptly reduced, back to its ambient level, on crossing the outer and inner toroidal (`periclinal') endodermis walls (also: Strasburger et al 1971). These two (sets of) transverse walls carry some ten `pores' each, and each pore is crossed by some $10^2$ plasmodesmata which apparently act as filter valves, diluting the incoming flux. The large number of these plasmodesmata, some $10^3$ per cell wall, may tell us that their mesoscopically small cross sections are essential for an efficient reduction of the osmolarity. We speculate that this reduction in $\pi$ is achieved by periodically squeezing the desmotubule - the connecting tube of a plasmodesma through the primary cell wall of the endodermis - via myosin VIII motors (Reichelt et al 1999) straining (torquing) an actin spiral that coils around it (Overall \& Blackman 1996; also: Olesen \& Robards 1990, Lucas et al 1993, Buchanan et al 2003, and Fig. 2e). The action of the myosin motors is thought to be triggered by Ca$^{++}$ ions (Berridge et al 1998). In this way, whilst the incoming ground water traverses the `sleeves', or `orifices' (of annular cross section) around the desmotubule, pure water is added to it downstream by periodically pressurizing the (filtering) plasmodesma.

   In more detail, a consensus has formed in recent years that the plasmodesmata are special cases of the endoplasmic reticulum (ER), a membrane-enclosed system of tubes between neighbouring cells which serves a supracellular exchange of material, or rather special cases of the vacuole, a subsystem of the ER which regulates the composition of the sap and/or the transport of pure water (Velikanov et al 2001). Whilst the cell fluid is called (cytoplasmic) `symplast', the fluid inside the ER is called `endoplast'. Ionic antiports (with H$^+$) keep the endoplast of a plasmodesma at a $10^4$ times higher Ca$^{++}$ concentration than its embedding symplast (Buchanan et al 2003, p.963). A vacuole can thus suck in pure water from its surroundings, and transport this water periodically across the cell wall whenever pinched off at its upstream end, and simultaneously pressurized beyond (like a pressurized bladder containing salt water, the prototype demonstration of a reverse osmosis). Note that backward diffusion of the diluted fluid is too slow to compete with the ordered incoming flow through the sleeve of the ER, so that an upward jump in $\pi$ is guaranteed.  

   Let us repeat and supplement the reasons of why we are convinced that the endodermis in the roothair zone can perform a reverse osmosis: To begin with, (i) the observed endodermis jump (in $\pi$ and $\psi$) requires a pump, and root pressure is well known to exist. If root pressure is to make the sap rise, it has to be exerted at their extremities (in order to push in the right direction), as is the case. (ii) The toroidal endodermis walls carry some $10^3$ plasmodesmata each which are thought to transfer pure water through their desmotubules, in addition to the concentrated ground water through their sleeves. (iii) The desmotubules of the plasmodesmata are embraced by an actin spiral which can cause peristaltic contractions (when torqued), strained by myosin VIII motors (Overall \& Blackman 1996, Reichelt et al 1999). (iv) The radial (`anticlinal') walls of the endodermis (and often of the exodermis as well, but of no other known walls) are strengthened by a ring of undulated, lignified Casparian strips which makes these cells rigid under varying pressures (Schreiber et al 1994). (v) The downstream ends of the desmotubules are surrounded by `neck constrictions' which have been proposed to act as `sphincters' (Olesen \& Robards 1990). These `actomyosin sphincters' may (likewise) be activated by a varying Ca$^{++}$ concentration, like the muscles of animals (Velikanov et al 2001). (vi) Root pressure - or exudation of excised roots - is inhibited by cytochalasin B and colchicine (Zholkevich et al 1990). (vii) Halophytes, with their stronger required reverse osmosis, have an extra set of pumps acting in series, in the exodermis.

   At what frequencies are the endodermis pumps expected to work? Symplastic flow velocities through the cortex - and hence likewise through the endodermis - are of order $v_c = 3 \mu$m/s. During half a pumping cycle, the enriched groundwater should have a chance to easily traverse the sleeve of a plasmodesma but only traverse a small fraction of the distance to the opposite cell wall, in order not to outrun the diluting pure water, hence flow through a distance $x$ between 0.1 and $3 \mu$m (cf. Fig. 2d,e), so that sinusoidal pumping would have to happen at frequencies $\nu$ in the interval
\begin{equation}
\nu = v_c/2x \in [0.5, 15]\,\mathrm{Hz} ,		
\end{equation}
not all that different from the frequency of the human heart. Can such a (high) pump frequency be maintained by the myosin VIII motors (which torque the actin spiral)? Myosin-motor molecules are observed to advance at speeds $\la 60 \mu$m/s, with steps of length $\la$nm, corresponding to step frequencies of several $10^4$/s (Reichelt et al 1999); our estimate therefore appears not to pose any problems. Note that muscles moving the human eye are known to be strained by myosin motors working on timescales of 10 ms, corresponding to frequencies of (even) $10^2$Hz (Thews et al 1989).

   Another quantity of interest is the number of pumps, or, equivalently, the number $N$ of endodermis cells in all the roothair zones of a big tree, for an assumed lifted mass rate $\dot M$ = ton/d of water to its crown. Using $A = 10^{3.5}\mu$m$^2$ for the typical outer toroidal wall area of an endodermis cell, and $v_c = 3 \mu$m/s for the typical traversal speed of the cortex (as above), we get
\begin{equation}
N = \dot M/\rho A v = 10^9 (\dot M/\mathrm{ton~per~day}) ,		
\end{equation}
i.e. $N = 10^9$ endodermis-cell traversals, or $10^{12}$ desmotubule traversals for a mass rate of one ton per day. Of course, smaller plants have fewer pumps, in direct proportion to their water demand $\dot M$.

   We thus mean to have shown that the rise of ground water in plants, both low plants and high plants, is likely owed to mechanical pumps in the roothair zone which perform a reverse osmosis, and whose excess pressure helps to propel the water columns in their xylem vessels.

\vspace{1cm}
{\Large\bf Acknowledgements}
\vspace{5mm}

Our cordial thanks go to Dieter Volkmann and Vadim Volkov for repeated help and encouragement throughout some ten years, as well as to Franti\u sek Balu\u ska, Heinz Felten, Fritz Lenz, Marko Marhl, and Gernot Thuma for more recent support. Andr\'e Bung kindly prepared our figures, and Ole Marggraf helped us with the online submission.       

\vspace{1cm}
{\Large\bf Appendix: The (preferred) Water Potential}
\vspace{5mm}

The motion of a body, or fluid in the gravity field of Earth is described by the equation of free fall:
\begin{equation}
\vec b = \vec g \; ,		
\end{equation}
i.e. its acceleration $\vec b$ is the free-fall acceleration $\vec g$ , also called `gravitational fieldstrength'; it points to the center of Earth. The vector $\vec g$ can be written as the (negative) gradient of the gravity potential $\phi_g \approx g z , \vec g = -\nabla \phi_g$, with $\nabla$ as the gradient operator. In a geographical map, the lines of constant height (isohypses) are lines of constant $\phi_g$; surface water flows downhill w.r.t. them.     

   If the body is a fluid and encaged in solid pipes (tubes), it can also flow uphill, namely in the direction of decreasing pressure $p$. In this case, the equation of motion (5) reads more generally:
\begin{equation}
\rho \vec b = \rho \vec g - \nabla p = -\nabla (\rho gz + p) .		
\end{equation}
Here the product $\rho \vec b$ is the force per volume acting on the fluid; it has been expressed as the (negative) gradient of the generalized pressure, $p + \rho gz$. If you are not (yet) familiar with the gradient vector, you may simply read it as the (ordinary) derivative along the filled tube: the derivative of the potential along the tube measures the force per volume exerted at each of its points.  

   In the vessels of the plants, we meet a yet more complicated situation, viz a spatially variable concentration of substance dissolved in the fluid (water). Dissolved matter attracts water, in proportion to the number density of its particles, independently of whether or not they are electrically neutral or charged, i.e. are molecules or ions. This `osmotic suction' can again be described by a potential, commonly denoted by $-\pi$, which plays the role of a negative pressure, the `osmotic pressure' $\pi$ ($\ge 0$); to first order in solute density $n_s$, $\pi$ equals the ideal-gas pressure of the solute: $\pi = n_s kT$. Equation (6) then generalizes to:
\begin{equation}
\rho \vec b = -\nabla(\rho gz + p - \pi) =: -\nabla \psi ,		
\end{equation}
wherein the linear combination $\rho gz + p - \pi$ is called $\psi$, the `water potential'. $\psi$ has the dimension of a pressure, force per area, like all its constituents.

   If, finally, the described system of tubes contains quite narrow ones, capillaries, like in seeds and spores, but also in leaves and buds and - to some extent - in every wood, one has to further add the `matric potential', or `imbibition potential' $\tau$, which describes the attractive force per area exerted on water by all the inner surfaces of a porous, hygroscopic (hydrophilic) body: 
\begin{equation}
\psi = \rho gz + p - \pi - \tau .		
\end{equation}
$\tau$ can be conveniently measured by bringing the porous body in contact with a (sucking) solution of suitable osmolarity: in (negative) pressure equilibrium, we have $\tau = \pi$. In extreme cases, both $\tau$ and $\pi$ can take gigantic values, of order kbar. 

   The water potential - as defined by us - has the property that a solution enclosed in a system of tubes (without pumps) flows always `downhill' in the generalized sense, i.e. towards smaller values of $\psi$. For example, osmotic and capillary suction can pull a column of sap upward in $z$, against gravity, at approximately constant hydrostatic pressure $p$, to heights approaching 10 km.

   As already mentioned above, $\psi$ has the dimension of a pressure, or energy density. It can be expressed through the chemical potential, or Gibbsian free enthalpy per particle $\mu$, as $\psi = \mu n$ , (with $n$ := number density). Water evaporates when the chemical potential of its vapour equals the chemical potential of the liquid. Consequently, the water potential of the gas phase (vapour) is some $10^3$ times smaller than that of the liquid. This factor of $10^3$ has mistakenly escaped from many textbooks on biology, with the (false) conclusion that transpiration was a gigantic force, reaching suctions as large as 1 kbar (Christian et al 2004, S. 86). Transpiration can be quite helpful, but only at moderate tensions.

\addtolength{\leftmargin}{7mm}
\setlength{\parindent}{-7mm}
\pagebreak

{\Large\bf References}

Berridge MJ, Bootman MD, Lipp P (1998) Calcium - a life and death signal. Nature  395:645-648

B\"ohm J (1893) Capillarit\"at und Saftsteigen. Ber. Deutsch. Bot. Ges. 11:203-212

Braune W, Leman A, Taubert H (1979) Pflanzenanatomisches Praktikum I. Gustav Fischer

Buchanan BB, Gruissem W, Jones RL (2003) Biochemistry \& Molecular Biology of Plants. ASPB Publications

Canny MJ (1995) A New Theory for the Ascent of Sap - Cohesion Supported by Tissue Pressure. Annals of Botany 75:343-357

Canny MJ (1998) Applications of the Compensating Pressure Theory of Water Transport. American Journal of Botany 85(7):897-909

Canny MJ (1999) The Forgotten Component of Plant Water Potential. Plant biol. 1:595-597

Christian A, Jaenicke J, Jungbauer W, K\"ahler H, Konopka HP, Kr\"uger D, M\"uller J, Paul A, Thomas F (2004) Biologie HEUTE entdecken, S II, Hg. Joachim Jaenicke \& Andreas Paul, Schroedel

Dixon HH, Joly J (1894) On the ascent of sap. Philosoph. Transact. of the Royal Soc.London, Ser. B 186:563-576

Hacke UG, Sperry JS (2003) Limits to xylem refilling under negative pressure in Laurus nobilis and Acer negundo. Plant, Cell and Environment 26:303-311

Hahn K (1993) Der Wassertransport in B\"aumen. Allgemeine Forstzeitschrift 22:1143-1150

Hales S (1726) Statical essays. Proceedings of the Royal Society of London, 1st edition

Koch GW, Sillett SC, Jennings GM, Davis S (2004) The limits to Tree Height. Nature 428:851-854

Kramer PJ, Kozlowski TT (1979) Physiology of Woody Plants. Academic Press

Kundt W (1998) The Hearts of the Plants. Current Science 75:98-102

Kundt W (2004) Astrophysics, A new Approach, Springer

Kundt W, Robnik M (1998) Water Pumps in Plant Roots. Russian Journal of Plant Physiology 45:262-269

L\"osch R (2001) Wasserhaushalt der Pflanzen. Quelle \& Meyer

Lucas WJ, Ding B, van der Schoot C (1993) Plasmodesmata and the supracellular nature of plants. New Phytologist 125:435-476

L\"uttge U, Higinbotham N (1979) Transport in Plants, Springer

L\"uttge U, Kluge M, Bauer G (1999) Botanik, 3. Aufl., Wiley-VCH

Matthews D (2006) The Water Cycle freshens up, Nature 439: 793-794

Milburn JA (1996) Sap Ascent in Vascular Plants: Challengers to the Cohesion Theory Ignore The Significance of Immature Xylem and the Recycling of M\"unch Water. Annals of Botany 78: 399-407

Mohr H, Schopfer P (1992) Lehrbuch der Pflanzenphysiologie. Springer-Verlag, Berlin

Morizet J, Robelin M (1972) La pouss\'ee radiculaire: D\'eterminations, M\'ecanisme, implications. Ann. Agron. 28 (5): 479-496

Nultsch W (1991) Allgemeine Botanik, 9. Aufl., Georg Thieme, Stuttgart

Olesen P, Robards AW (1990) The Neck Region of Plasmodesmata. In: Robards AW, Lucas WJ, Piits JD, Jongsma HJ, Spray DC (eds) Parallels in Cell to Cell Junctions in Plants and Animals. Springer, pp. 145-170

Overall RL, Blackman LM (1996) A Model of the Macromolecular Structure of Plasmodesmata. Trends in plant science 1, No.9:307-311

Pfanz H, Aschan G, Langenfeld-Heyser R., Wittmann C, Loose M (2002) Ecology and ecophysiology of tree stems: corticular and wood photosynthesis. Naturwissenschaften 89:147-162

Pfeffer W (1881) Pflanzenphysiologie. Ein Handbuch des Stoffwechsels und Kraftwechsels in der Pflanze. Erster Band: Stoffwechsel. Verlag Wilhelm Engelmann, Leipzig

Pickard WF (1981) The Ascent of Sap in Plants. Prog. Biophys. molec. Biol. 37:181-229

Preston RD (1952) Movement of water in higher plants. In: Frey-Wyssling A, ed. Deformation and Flow in biological systems. North Holland Publishing Co., Amsterdam, 257-321

Read D (1997) The ties that bind. Nature 388:517-518

Reichelt S, Knight AE, Hodge TP, Baluska F, Samaj J, Volkmann D, Kendrick-Jones J (1999) Characterization and localization of the unconventional myosin VIII in plant cells and its localization at the post-cytokinetic cell wall. Plant Journal 19: 555-568

Sachs J (1887) Vorlesung \"uber Pflanzen-Physiologie. 2. Aufl., Verlag Wilhelm Engelmann, Leipzig

Salisbury FB, Ross CW (1992) Plant Physiology. Wadsworth Publ. Company, Belmont

Schreiber L, Breiner HW, Riederer M, D\"upggelin M, Guggenheim R (1994) The Casparian Strip of Clivia miniata Reg. Roots: isolation, Fine Structure and Chemical Nature. Bot. Acta 107:353-361

Schwenke H, Wagner E (1992) A new concept of root exudation. Plant, Cell and Environment 15:289-299

Steudle E (1995) Trees under Tension. Nature 378:663-664

Steudle E (2001) The Cohesion-Tension Mechanism and the Aquisition of Water by Plant Roots. Annu. Rev. Plant Physiol. Plant Mol. Biol. 52:847-875

Strasburger E, Noll F, Schenck H, Schimper AFW, v. Denffer D, Schumacher W, M\"agdefrau K, Ehrendorfer F (1971) Lehrbuch der Botanik. Gustav Fisher Verlag, Stuttgart

Thews G, Mutschler E, Vaupel P (1989) Anatomie, Physiologie,, Pathophysiologie des Menschen. Wissenschaftliche Verlagsgesellschaft Stuttgart, 3. Aufl.; S. 406

Tyree MT (1997) The Cohesion-Tension theory of sap ascent: current controversies. Journal of Experimental Botany 48:1753-1765

Tyree MT (2003) The ascent of water. Nature 423:923

Tyree MT, Zimmermann MH (2002) Xylem Structure and the Ascent of Sap, Springer

Ursprung A, Blum G (1921) Zur Kenntnis der Saugkraft IV: Die Absorptionszone der Wurzel. Der Endodermissprung. Berichte der deutschen botanischen Gesellschaft 39:70-79

Velikanov GA, Volbueva OV, Khokhlova LP (2001) The Study of the Hydraulic Conductivity of the Plasmodesmal Transport Channels  by the Pulse NMR Method. Russian Journal of Plant Physiology 48:318-325

White PR, Schuler E, Kern JR, Fuller FH (1958) ``Root Pressure'' in Gymnosperms. Science 128:308-309

Woodward I (2004) Tall storeys. Nature 428:807-808

Zholkevich VN (1991) Root Pressure. In: Waisel Y, Eshel A, Kafkafi U (eds.) Plant Roots, The Hidden Half. Marcel Dekker, New York

Zholkevich VN, Chugunova TV, Korolev AV, Timiriazev KA (1990) New Data on the nature of root pressure. Studia Biophysica 136:209-216

Zimmermann U, Haase A, Langbein D, Meinzer F (1993) Mechanisms of long-distance water transport in plants: a re-examination of some paradigms in the light of new evidence. Phil. Trans. R. Soc. London B 341:19-31

Zimmermann U, Rygol J, Balling A, Kl\"ock G, Metzler A, Haase A (1992) Radial Turgor and Osmotic Pressure Profiles in Intact and Excised Roots of Aster tripolium. Plant Physiology 99:186-196

Zimmermann U, Schneider H, Wegner L H, Haase A (2004) Water ascent in tall trees: does evolution of land plants rely on a highly metastable state?, New Phytologist 162: 575-615

\addtolength{\leftmargin}{-7mm}
\setlength{\parindent}{7mm}
\pagebreak

{\Large\bf Figure Captions}
\vspace{5mm}

Fig. 1: Schematic drawing of how a water column can rise in a vertical tube, pulled, a) by surface tension inside a narrow tube of hydrophilic material - to a height of $h = 15$ m ($\mu$m/$r$) for a tube radius of $r$ in units of $\mu$m  -  or b) by osmotic suction - to a height of $h = 10$ m ($\pi$/bar) for an osmotic pressure $\pi$ in units of bar - the weight of the column being carried by semi-permeable separating `floors' which plants realize as sloping cell walls with bordered pits, or perforation plates.

\vspace{5mm}

Fig. 2: From the Tree to the Plasmodesma, in four steps of enlargement: Tree (a) $\to$ Root Segment (b) $\to$ Root Tip (c) $\to$ Endodermis and Pericycle (d) $\to$ Plasmodesma (e). The factors of enlargement are {100, 32, 32, 100} respectively; the small white squares mark the areas of enlargement.

  The ground water (red) is sucked in osmotically through the root hairs and cortex (c), then diluted again to soil-like osmolarities by active injection of pure water through the pasmodesmata (e) on entering the endodermis and pericycle cells (d). Marked in green (in a, c) are phloem vessels in which the products of photosynthesis are transported towards the zones of growth: branches, fruits, and roots.

  Figure c shows partially a section, partially a view from outside; the endodermis jump (in osmolarity) takes place in transition from the cortex to the central cylinder, starting in the endodermis layer (tinted yellow). In Fig. d, the radial cell walls of the endodermis layer form a corrugated, lignified girdle of Caspary strips (tinted purple) which enhances their rigidity. Double arrows stand symbolically for `pores' in the cell wall, each of which is traversed by some 100 plasmodesmata. 

  Figure e is only resolved by the electron microscope. Shown is a section of one `pore' with two neighbouring plasmodesmata in which membrane-enclosed tubes - an ER, or rather a vacuole (Velikanov et al 2001) - traverse the primary cell wall. The lower plasmodesma is shown in the transmission phase, the upper one in the locked-up phase. Whilst the sleeve domain allows the passage of the incoming flow (of high osmolarity; red arrows), the central desmotuble takes care of sluicing in pure water (blue arrows). The desmotuble is embraced by an actin spiral (green) which can be tightened by myosin VIII motors, such that pure water is squeezed out in the forward direction, towards the pericycle. Membrane segments which are drawn-in broken map earlier snapshots. The `sphincter' sketched in ochre at the downstream end can probably close the valve whenever the root pressure is to be reduced (Olesen \& Robards 1990).    

\vspace{5mm}

Fig. 3: Guttation shown for the herb `Frauenmantel' (= Lady's Mantle Rose): The leaves bottom left stand without roots in a glass of water, contrary to those top right which grow in soil, at whose leaf edges can be recognised droplets of guttation. The exposure was made at 7 o´clock in the morning when all leaves still carried large numbers of dew droplets.

\vspace{5mm}

Fig. 4: Guttation shown for stalks of Grass: The stalks at the left stand in water, contrary to the stalks (with roots) at the right. Only the stalks with roots show guttation. 

\vspace{5mm}

Fig. 5. The Roothair Zone of germinating Cress: is distinctly recognisable for several-days old germinating cress in a Petri bowl.

\begin{figure}
\includegraphics[width=15cm]{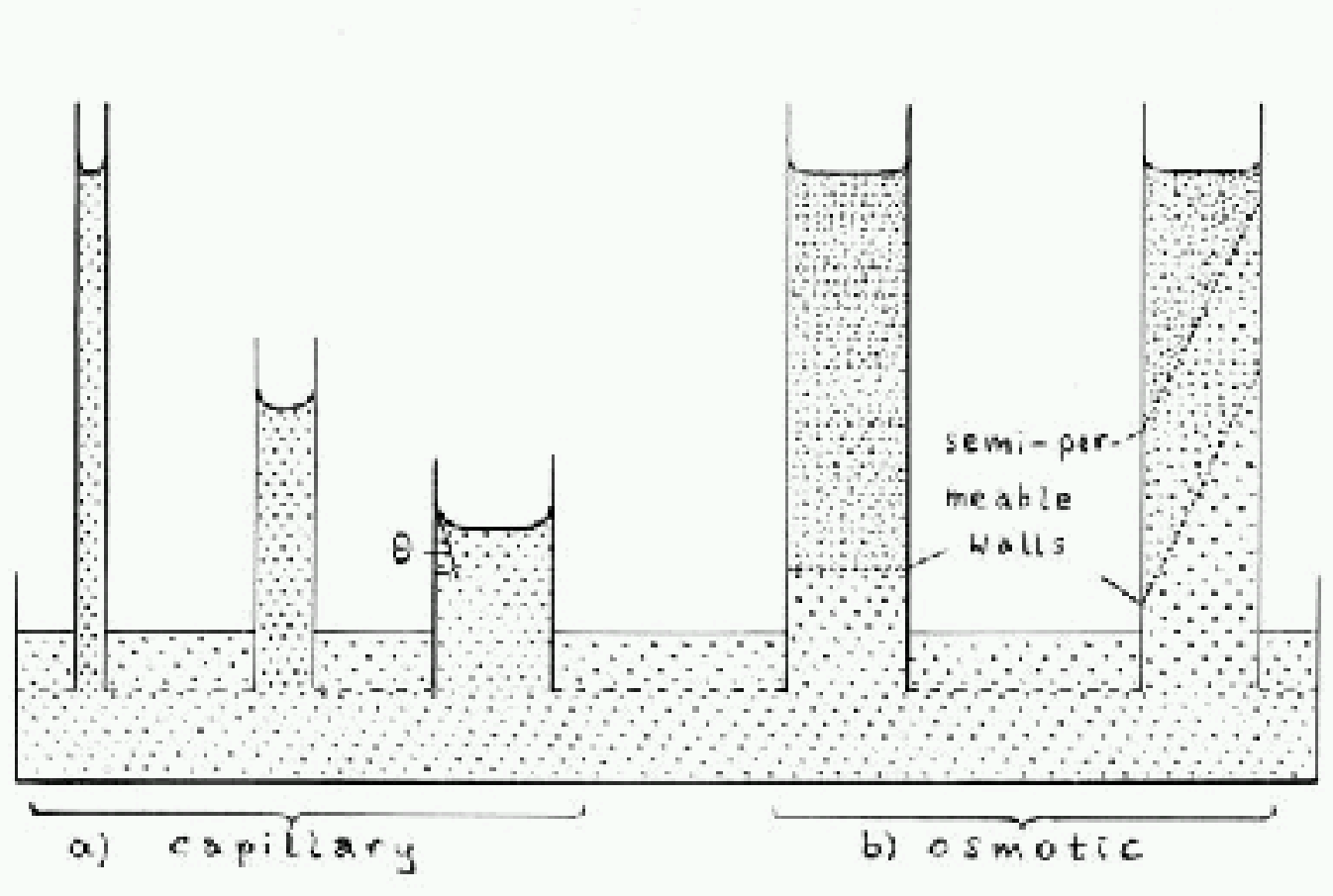}
\caption{}
\end{figure}

\begin{figure}
\includegraphics[width=15cm]{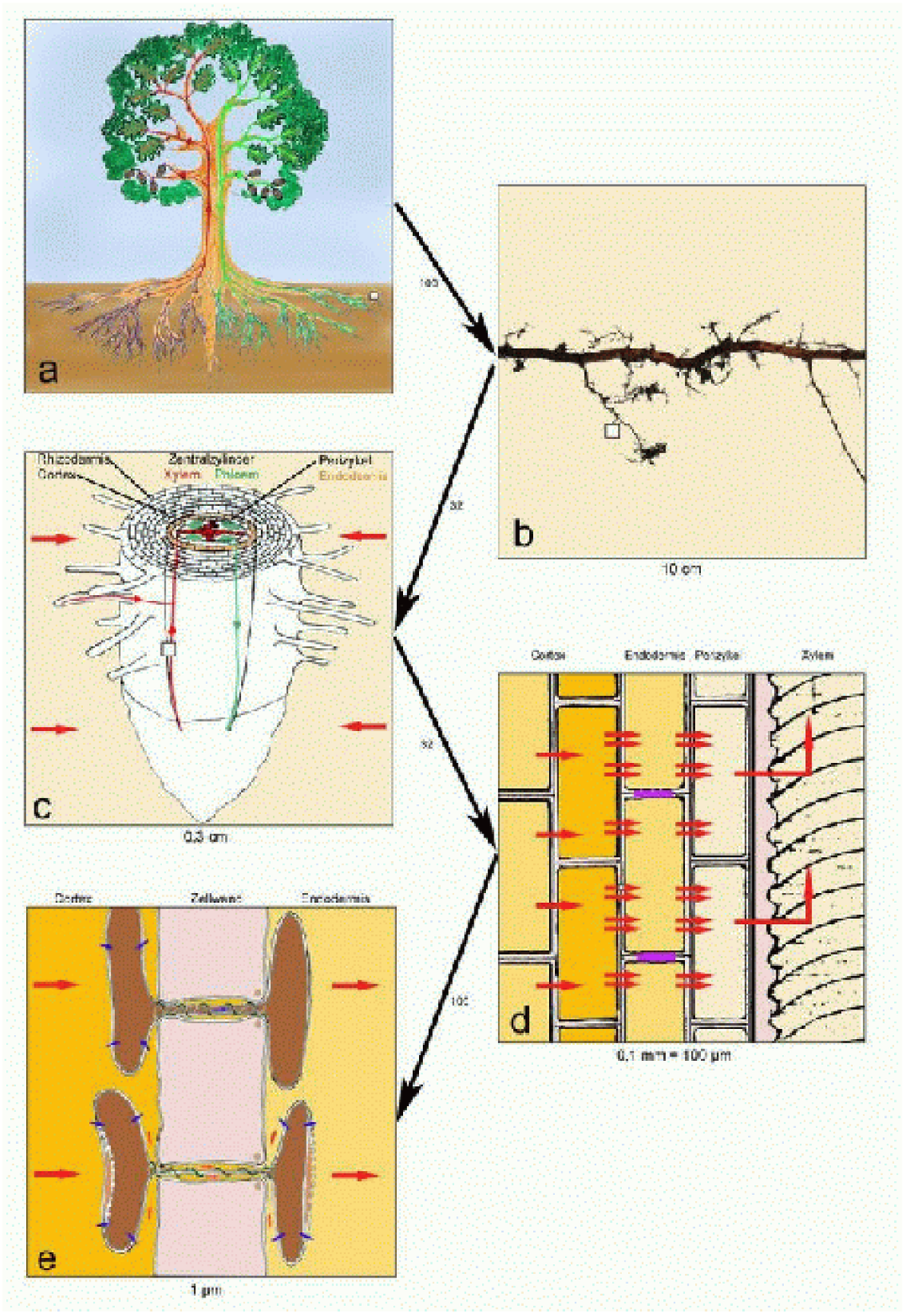}
\caption{}
\end{figure}

\begin{figure}
\includegraphics[width=15cm]{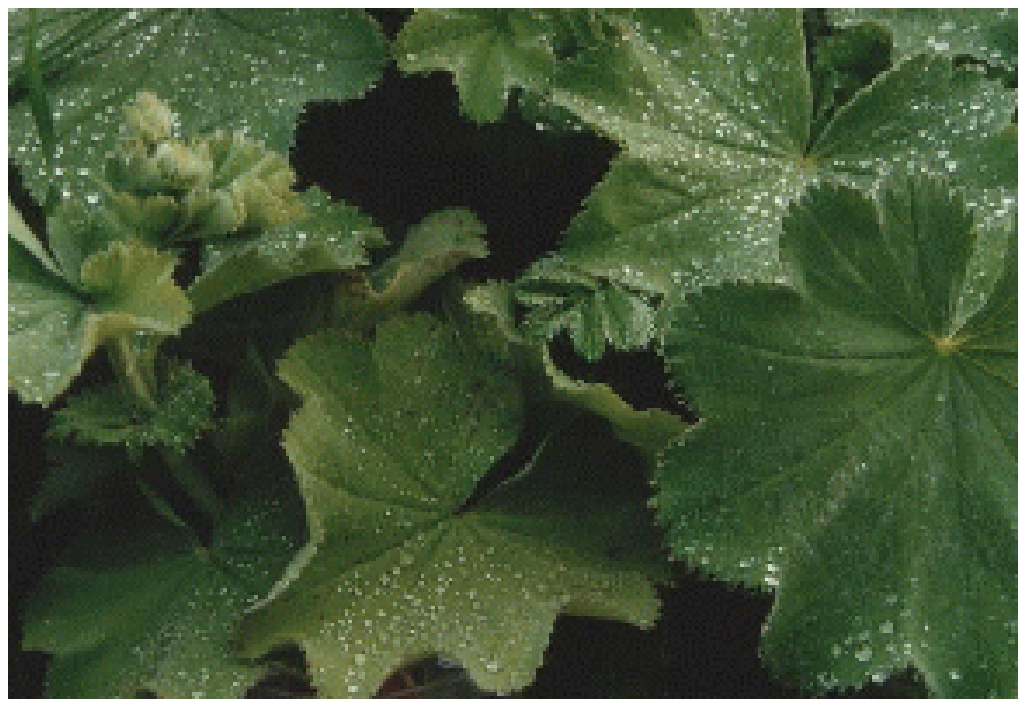}
\caption{}
\end{figure}

\begin{figure}
\includegraphics[width=15cm]{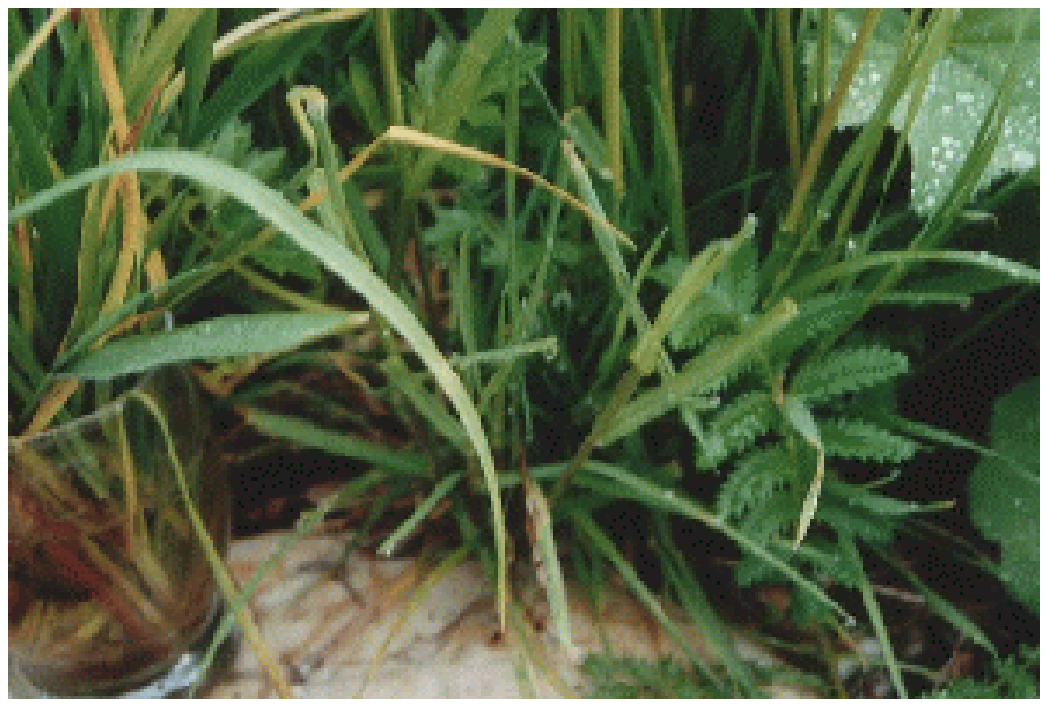}
\caption{}
\end{figure}

\begin{figure}
\includegraphics[width=15cm]{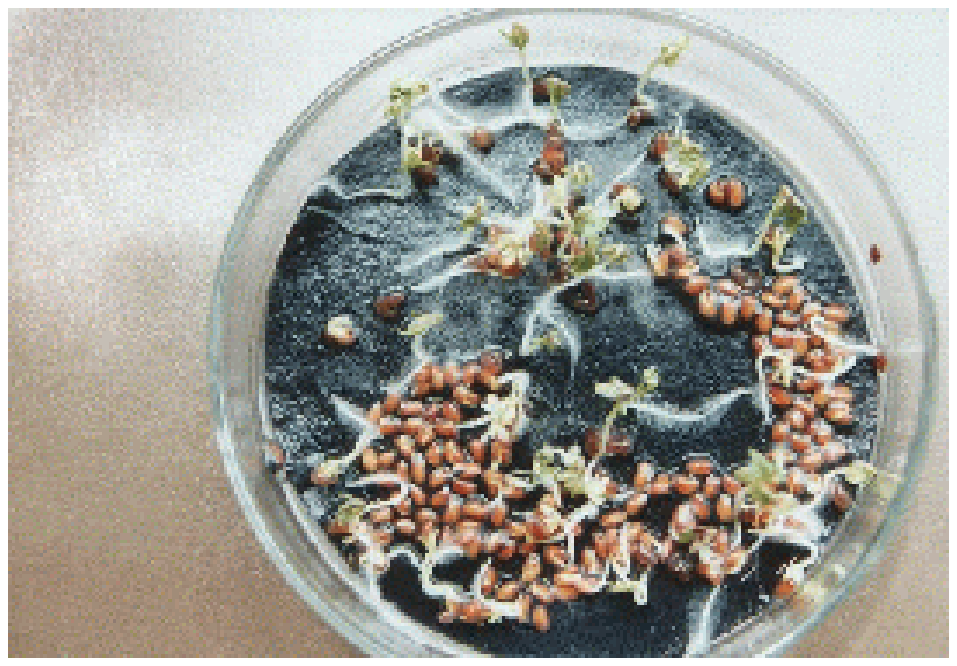}
\caption{}
\end{figure}

\end{document}